# Silicon photonic paper-clip spiral delay lines with ultra-low delay loss of 0.5 dB/ns

BATOUL HASHEMI,[1, †] MANUEL ARTURO MÉNDEZ-ROSALES,[1, †] PARIMAL EDKE, [1] MOHAMMAD REZAUL ISLAM,[1] RANJAN DAS,[1] ANDREW P. KNIGHTS,[1] AND JONATHAN D. B. BRADLEY[1, *]

[1] *Department of Engineering Physics, McMaster University, 1280 Main Street West, Hamilton, ON L8S 4L7, Canada*
*[hashemb@mcmaster.ca](hashemb@mcmaster.ca)*

**In this work, we demonstrate compact paper-clip spiral silicon photonic waveguides with ultra-low delay loss. We characterize the optical loss and group delay of single-mode and multi-mode silicon waveguides across the telecom O-, S-, C-, and L-bands. For spiral devices with 2.0-μm-wide waveguides, we measure propagation losses of 0.11 and 0.06 dB/cm at 1310 and 1550 nm, representing 10- and 20-times improvements, respectively, compared to the single-mode waveguides. Additionally, we observe a group delay of 1163 ps for a 9.5 cm-long waveguide with a compact device footprint of (0.30 × 3.00) mm$^2$, yielding a delay loss of 0.5 dB/ns. These results are highly promising for large-scale silicon photonic integration, delay lines, and on-chip programmable systems.**

Silicon-on-insulator (SOI) has emerged as a leading platform for integrated photonic technologies, offering key advantages like compatibility with complementary-metal-oxide-semiconductor (CMOS) processes, mature fabrication processes, high integration density, low cost, and strong optical confinement enabled by its high refractive index contrast [1–3]. These advantages make SOI one of the primary platforms for emerging research areas such as neural networks [3], quantum photonics [4], microwave photonics [5,6], and high-speed communications [7]. Waveguides are essential building blocks in silicon photonics, forming the basis for active and passive devices [1,3], and functioning as standalone devices or interconnects in complex photonic integrated circuits (PICs), for applications such as true-time delay [8–12], high-Q cavities [13–15], and beamforming tomography [16]. Minimizing propagation loss in waveguides remains a critical priority for meeting system power budgets, enhancing device performance, and maximizing signal to noise ratio. Significant efforts have been devoted to achieving low-loss optical waveguides using CMOS-compatible platforms like silicon dioxide (SiO$_2$), silicon nitride (Si$_3$N$_4$), and silicon. SiO$_2$ and Si$_3$N$_4$ waveguides can achieve extremely low propagation losses, down to ~0.1 dB/m [1,17]. However, their weak confinement necessitates large bend radii, limiting integration density. In contrast, silicon waveguides allow for a compact footprint, but inherently suffer from increased scattering losses [18]. Propagation loss in silicon waveguides arises from interface scattering, substrate leakage, absorption, and bend radiation losses, with scattering being the dominant factor for losses [18,19]. Two main strategies have been explored to reduce propagation losses. The first involves fabrication process optimization, including chemical mechanical polishing [12,20], oxidation-based etchless processes [21,22], wet chemical etching [23], hydrogen plasma treatments [24], and anisotropic etching [25]. While effective, these techniques often require post-processing steps that may not be compatible with standard foundry workflows. The second approach focuses on waveguide design to minimize modal interaction with sidewalls, for example, by using ultra-thin or shallow-ridge waveguide structures [26,27], thick SOI platforms [28], or wider waveguides that reduce sidewall overlap [19,29–31]. Curvature-induced radiation losses and multi-mode crosstalk in wide waveguides can be minimized by designing adiabatic waveguide transitions based on analytic curves (such as clothoid or sinusoidal curves), enabling smooth transitions [31]. Using these approaches, propagation losses as low as 0.13 dB/cm at 1550 nm [30], resonators with quality factors exceeding 10$^7$ [13], delay ranges up to 12.7 ns with a high delay-density ratio of 3299 ps/mm$^2$ [29], and 5.1 ns delay range with delay loss of 2.3 dB/ns [19] have been recently reported.

In this work, we demonstrate ultra-low-loss silicon paper-clip spiral waveguides fabricated through a standard multi-project wafer (MPW) run at the Advanced Micro Foundry (AMF) in Singapore. By incorporating Euler bends and extended multi-mode straight segments, bend-induced losses are reduced while maintaining a compact device footprint. We observe propagation losses of 0.11 ± 0.03 dB/cm and 0.06 ± 0.03 dB/cm at 1310 and 1550 nm, respectively, for (2.0 × 0.22) μm$^2$ silicon spiral waveguides. Group delay measurements further demonstrate a maximum delay of 1163 ± 30 ps with a delay loss of 0.5 ± 0.2 dB/ns. These results demonstrate a compact spiral geometry that pushes the limits of waveguide performance on the SOI platform and is highly attractive for large-scale silicon photonic integration.

Figure 1(a) shows a schematic of the silicon photonic paper-clip spiral waveguide. The design leverages long straight waveguides, minimizing bend-associated losses and inter-modal crosstalk [19,31]. Waveguide widths of 0.5, 1.0, 1.5, and 2.0 µm were chosen, with 2.0 µm exhibiting the lowest propagation loss [19]. Following a similar approach as in [13,14], the design incorporates $180^0$ Euler bends to minimize mode-mismatch and bend-induced losses. The design further allows for convenient layout stacking for PIC applications requiring multiple spirals. The waveguide spacing is ≥5 µm between sidewalls to prevent optical coupling between adjacent waveguide paths. Edge couplers with 0.18-µm-wide facet-tip taper linearly to single-mode waveguides over 75 µm at the input and output ports. A smooth, low-loss transition between waveguide cross-sections is achieved by adiabatically tapering the waveguide width from the single-mode waveguide to the target multi-mode waveguide over a 400-µm length. Eigenmode expansion simulations indicate that for the widest width transition (0.5- to 2-µm), a 100-µm-long taper yields a multi-mode crosstalk of ≤–40 dB. Each $180^0$ bend is formed by two mirrored Euler curve segments, symmetric about the mid-point. The bending radius along the device arclength is kept at ≥ 30 µm, with the minimum bend radius located at the midpoint of the two innermost 180° Euler bends. The sinusoidal S-bends feature an adiabatic and continuous curvature transition, with a minimum bend radius exceeding 100-µm, and are antisymmetric about the mid-point. Mode mismatch losses at straight-to-bend transitions are minimized by designing bends with zero-curvature endpoints, ensuring curvature continuity throughout the device. Like the waveguide-width tapers, higher-order mode excitation is suppressed by adiabatic curvature transitions, defined here by the rate of change of curvature along the arclength. Owing to the elongated aspect ratio of the design, the central S-bends have a slower curvature variation than the 180° Euler bends, where the innermost bends dominate the overall curvature profile. Finite-difference time-domain simulations were used to extract transmission coefficients for the first five TE modes showing crosstalk level of ≤–30 dB between the fundamental TE mode and higher order modes across the C-band.

Figures 1(b)-(e) show the fundamental TE mode at 1550 nm for each waveguide width simulated using a finite-difference eigenmode (FDE) solver. The group index was also obtained at 1310 and 1550 nm using the FDE solver. The delay per unit length $\Delta\tau$ (ps/cm) was calculated using $\Delta\tau = (n_g/c_0) \cdot L/L_0$, where $n_g$ is the group index, $c_0$ is the speed of light in a vacuum, $L$ is spiral length (cm), and $L_0$ is a 1-cm reference length. The resulting values are summarized in Table 1. The difference in delay per unit length between a straight and a bent waveguide is largest for the widest waveguide, calculated as $1 \times 10^{-6}$ ps/cm for a 30 µm bend radius, which is considered negligible. Therefore, given the design of the paper-clip waveguide path, curvature-related variations in delay per unit length are also considered insignificant. Figure 1(f) shows a micrograph of paper-clip waveguides including a set of spirals with lengths of 9.5, 5.1, 2.9, and 1.2, (from top to bottom) and corresponding device footprints of (0.30 × 3.00) mm², (0.19 × 3.00) mm², (0.13 × 3.00) mm², and (0.09 × 3.00) mm², respectively.

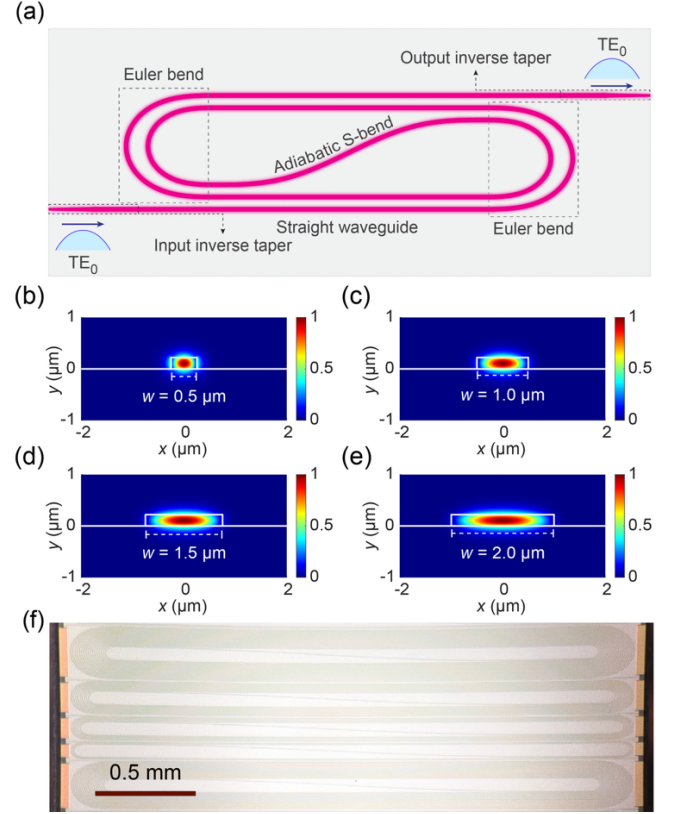

Fig. 1. (a) Paper-clip spiral waveguide schematic with labelled constituent curve segments. Simulated transverse electric field ($E_y$) fundamental mode profiles at 1550 nm for waveguide geometries of (b) 0.5 × 0.22 µm², (c) 1.0 × 0.22 µm², (d) 1.5 × 0.22 µm², (e) 1.0 × 0.22 µm². (f) Optical micrograph of fabricated paper-clip waveguides.

Table 1. Simulated group index and delay per unit length for single-mode and multi-mode waveguides

| Waveguide width (µm) | 1310 nm | | 1550 nm | |
|---|---|---|---|---|
| | $n_g$ (RIU) | $\Delta\tau$ (ps/cm) | $n_g$ (RIU) | $\Delta\tau$ (ps/cm) |
| 0.5 | 4.162 | 138.8 | 4.241 | 141.5 |
| 1.0 | 3.906 | 130.3 | 3.880 | 129.4 |
| 1.5 | 3.858 | 128.7 | 3.808 | 127.0 |
| 2.0 | 3.841 | 128.1 | 3.782 | 126.2 |

To characterize the waveguide transmission, we used tunable laser sources covering the 1240–1380 nm and 1450–1650 nm wavelength ranges. Three-paddle polarization controllers were added after the tunable laser source to ensure excitation of the fundamental TE mode. The average propagation loss in the spirals at each wavelength was obtained using the cutback method, with measurements performed across four chips with the same design. We determined the propagation losses for waveguide widths of 0.5, 1.0, 1.5, and 2.0 μm; the corresponding averages and standard deviations are plotted in Fig. 2. The propagation losses at 1310 nm are 1.10 ± 0.1, 0.40 ± 0.03, 0.14 ± 0.01, and 0.11 ± 0.03 dB/cm, for waveguide widths of 0.5, 1.0, 1.5, and 2.0 μm, respectively. An increase in propagation loss near 1380 nm may be attributed to OH absorption in the $SiO_2$ cladding [32]. At 1550 nm, the corresponding losses are 1.30 ± 0.10, 0.30 ± 0.04, 0.12 ± 0.06, and 0.06 ± 0.03 dB/cm for the same set of waveguide widths. The fiber-chip coupling loss is similar across all widths and, for the 2.0-μm-wide waveguide sets, is 0.80 ± 0.10 dB/facet at 1310 nm and 0.45 ± 0.05 dB/facet at 1550 nm. The insets in Fig. 2(a) and Fig. 2(b) show the measured propagation loss for the 2.0-μm-wide waveguides in the O-, S-, C- and L-bands across the four chips, confirming consistent performance. These results compare favorably with other ultra-low-loss Si waveguide demonstrations [19,29,30].

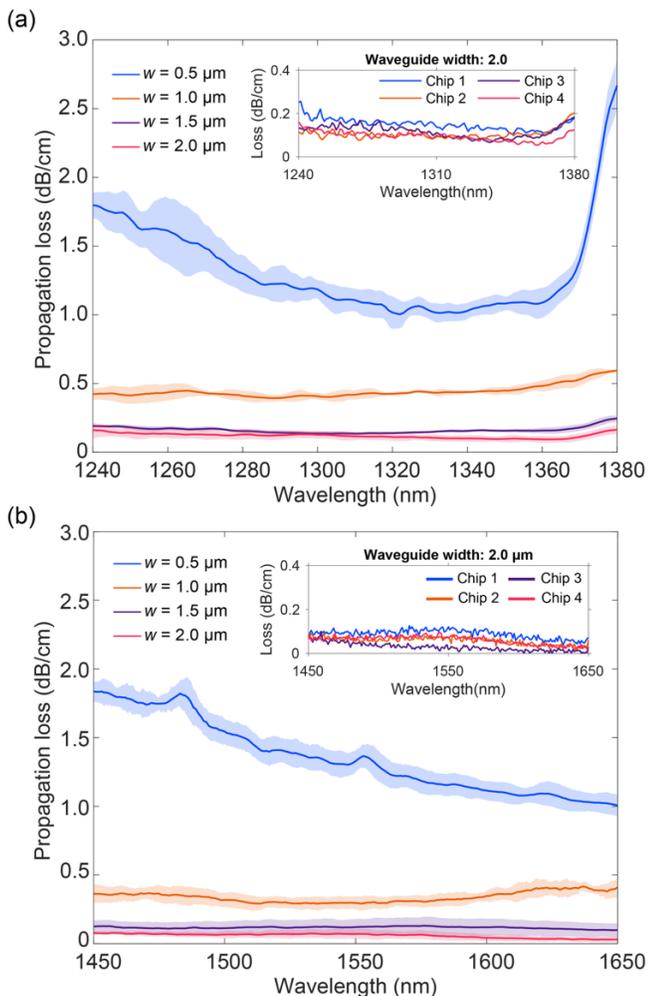

Fig. 2. Propagation loss versus wavelength for the (a) O-band (1240–1380 nm) and (b) S-, C- and L-bands (1450–1650 nm) for waveguide widths of 0.5, 1.0, 1.5, and 2.0 μm. Insets show the measured propagation loss across four chips for the 2.0-μm-wide waveguides.

We measured the optical group delay characteristics using the methodology described in [33] and the experimental setup shown in Fig. 3(a). In this approach, an optical carrier from a tunable continuous-wave (CW) laser source is modulated using an electro-optic (EO) Mach-Zehnder modulator with a radio frequency (RF) signal supplied by an electrical vector network analyzer (EVNA). Polarization controllers (PC) are placed before and after the modulator to ensure fundamental TE mode transmission through the EO modulator and coupling at the chip facet. Micro-positioner XYZ translation stages are used to align the tapered fibers to the edge coupler input and output spiral ports. A photodetector (PD) and RF amplifier extract the electrical signal, which is sent back to the EVNA. This approach enables the precise and indirect determination of optical group delay from the phase information of an $S_{21}$ EVNA measurement, representative of the transmission of the RF signal through the device, by measuring the RF phase shift with respect to the input signal phase ($\varphi_{21}$). Delay measurements were performed at 1550 nm, with a 5 GHz RF signal over a 20 MHz span. Figure 3(b) presents the measured loss as a function of delay.

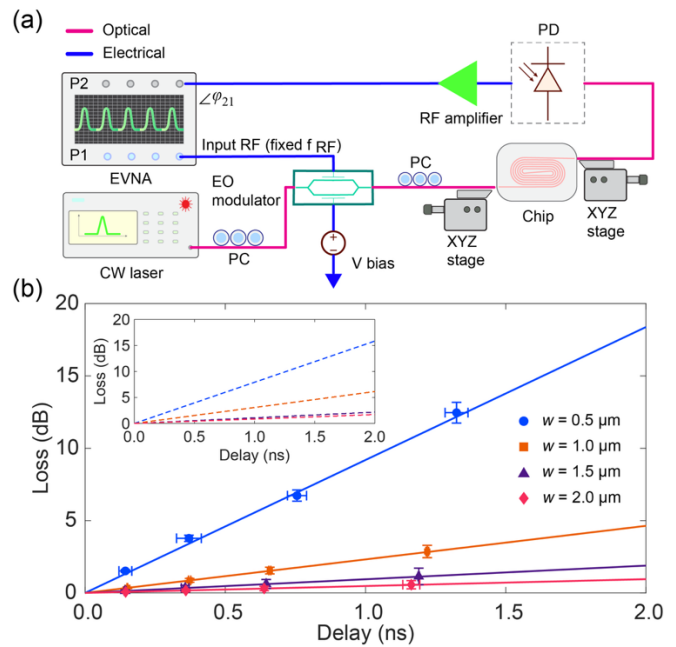

Fig. 3. (a) Optical group delay measurement setup and (b) Measured propagation loss vs. optical group delay. Data points represent the measured group delay and propagation loss for each waveguide width and length, while the solid lines show the calculated delay based on the measured propagation loss and simulated group index at 1550 nm. The inset displays the expected delay derived from the measured propagation loss and simulated group index at 1310 nm.

The data points represent the measured values, and the error bars indicate the standard deviation of delay and loss across four chips. The solid lines are calculated based on the simulated group index and the measured propagation loss. The total loss is related to the group delay by the equation $a = \alpha_{\text{prop}} \cdot L = \alpha_{\text{prop}} \cdot \left(\frac{c_0 \Delta \tau}{n_g}\right)$, where $a$ is the total loss in (dB), and $\alpha_{\text{prop}}$ is the propagation loss in (dB/cm). for the 9.5-cm-long paper-clip waveguide at 1550 nm, the 0.5-µm-wide waveguide has a delay loss of 9.2 ± 0.7 dB/ns with a net group delay of 1324 ± 40 ps, while the 2-µm-wide waveguide has a delay loss of 0.5 ± 0.2 dB/ns with a net group delay of 1163 ± 30 ps. Using the measured propagation loss and the group index values listed in Table 1, the estimated delay losses at 1310 nm are 7.9 dB/ns and 0.9 dB/ns for the 0.5- and 2.0-µm-wide waveguides, corresponding to calculated net group delays of 1315 and 1213 ps, respectively, as shown in the inset of Fig. 3(b). The measured delay loss at 1550 nm is lower than that in other SOI platforms and comparable to SiN. The delay density lies between the values reported for other SOI and SiN platforms and can be improved through spiral layout optimization [19,29,34].

In summary, we have designed, fabricated, and characterized paper-clip spiral silicon waveguides with ultra-low delay loss. For the waveguide width of 2-µm, we obtained the propagation loss of 0.11 ± 0.03 and 0.06 ± 0.03 dB/cm at 1310 and 1550 nm, respectively and achieved a group delay of 1163 ± 30 ps for a 9.5-cm-long spiral, corresponding to a delay loss of 0.5 ± 0.2 dB/ns. The improved performance is attributed to reduced optical mode interaction with the waveguide sidewall. The paper-clip design leverages long straight waveguide sections that constitute most of the device length, effectively minimizing radiative bend losses and sidewall scattering. Further optimization of the spiral structure could improve device performance and footprint. This work highlights the advantage of multi-mode waveguides for applications requiring low delay loss characteristics, employing a compact, curvature-optimized paper-clip structure to achieve high net group delay.

**Funding.** This research was funded by the Canada Foundation for Innovation; High-throughput and Secure Networks Challenge program; MacDonald, Dettwiler and Associates (MDA); Mitacs; National Research Council Canada; Natural Sciences and Engineering Research Council of Canada.

**Acknowledgment.** We thank CMC Microsystems for facilitating the silicon photonic design and fabrication and Dr. Natalia Nikolova for valuable discussions.

**Disclosures.** The authors declare no conflicts of interest.

**Data availability.** Data underlying the results presented in this paper are not publicly available at this time but may be obtained from the authors upon reasonable request.